# Data Management for High-Throughput Genomics

Uwe Röhm
University of Sydney
Sydney, Australia
roehm@it.usyd.edu.au

José A. Blakeley
Microsoft Corporation
Redmond, USA
joseb@microsoft.com

## ABSTRACT

Today's sequencing technology allows sequencing an individual genome within a few weeks for a fraction of the costs of the original Human Genome project. Genomics labs are faced with dozens of TB of data per week that have to be automatically processed and made available to scientists for further analysis. This paper explores the potential and the limitations of using relational database systems as the data processing platform for high-throughput genomics. In particular, we are interested in the storage management for high-throughput sequence data and in leveraging SQL and user-defined functions for data analysis inside a database system. We give an overview of a database design for high-throughput genomics, how we used a SQL Server database in some unconventional ways to prototype this scenario, and we will discuss some initial findings about the scalability and performance of such a more database-centric approach.

## 1. INTRODUCTION

Genetic analysis labs are facing a serious data management problem: The development of high-throughput gene sequencing instruments makes mass genomics feasible, but at the same time produces gigabytes of sequencing data per experiment that need to be stored, aligned and analyzed. While it took 10 years and $3B dollars to produce a first draft of the human reference genome (approx. 3.5 billion base pairs (bp)), the current generation of sequencing instruments is able to sequence between 2 to 4 billion bases in only a few days [8][19].

The current approach to data management for high-throughput genomics is file-centric and involves a large number of separate files containing a text-format or a proprietary binary format. Most of these formats are insufficiently documented and do not include meta-data. Some file formats that include metadata (e.g., HDF5) do not separate the data's representation from its conceptual data model. This makes data workflow management very complicated and the analysis becomes inefficient. The current file-centric approach simply does not scale to the terabyte needs of high-throughput genomics.

On the other hand, traditional database technology also has shortcomings that prevent it from becoming commonly used in this scientific domain: Database systems are optimized for fast processing of small and precise business data records. Everything beyond these core assumptions is a challenge for current database engines. This includes large BLOB-style attributes – such as gene sequences –, uncertain data – such as most scientific data –, computational intensive data processing functions – such as sequence alignment algorithms –, and efficient support for data provenance and annotation management. But perhaps most importantly, database systems are not easy to deploy and use by the average scientist. They not only require some specific technical skills, but also a certain way of thinking about data design and declarative data access which apparently is not widely embraced by the scientific community.

A common perception in the Bioinformatics community is that "*Traditional DBMSs require significant performance tuning for scientific applications, a task that scientists are neither well prepared for, nor eager to address. [...] Most importantly, the complex data structures and data types needed to represent the data and relationships are not supported well in most databases*"[15]

In this paper, we are interested to find out whether this perception is true. We study two different scenarios from high-throughput genomics, the 1000 Genomes project [1] and digital gene expression analysis [20], and investigate how current database technology can be used for efficient data management. In particular, we are interested in leveraging the extensibility and file streaming features of the latest SQL Server 2008 for processing of gene sequence data [5][10].

Our findings show that there are shortcomings in both approaches: Existing file-centric efforts for genomics violate all best practices of the database community, from basically non-existing data models to strong data-program dependencies. At the same time, current DBMSs lack optimizations for complex user-defined functions and in general do not support the right abstractions for probabilistic data or data provenance. Nevertheless, we show how efficient genomic data management can be achieved with today's extensible database systems and which trade-offs between strict data modeling and external tool support have to be made.

## 2. BACKGROUND

Let us first have a look at the current state-of-the-art in high-throughput genomics and in extensible database technology.

### 2.1 Gene Sequencing Workflows

Gene sequencing workflows consist of five phases (cf. Figure 1):

**Phase -1: Sample Preparation**
Using sophisticated chemical processes, a sequencing lab first prepares several so-called libraries of DNA/mRNA samples by chopping and replicating the original nucleotide sequences into fragments that are short enough to be processed by the current sequencer machines (the current technology limit is about 300bp). Each library is then prepared onto a flowcell (or slide) that consists of 8 physical lanes. One lane is normally reserved for a control sample, while the remaining seven contain the material to be sequenced. Each lane is logically divided into about 300 tiles which are the actual unit-of-measurement of the sequencers.

**Figure 1 Typical Gene Sequencing Pipeline**

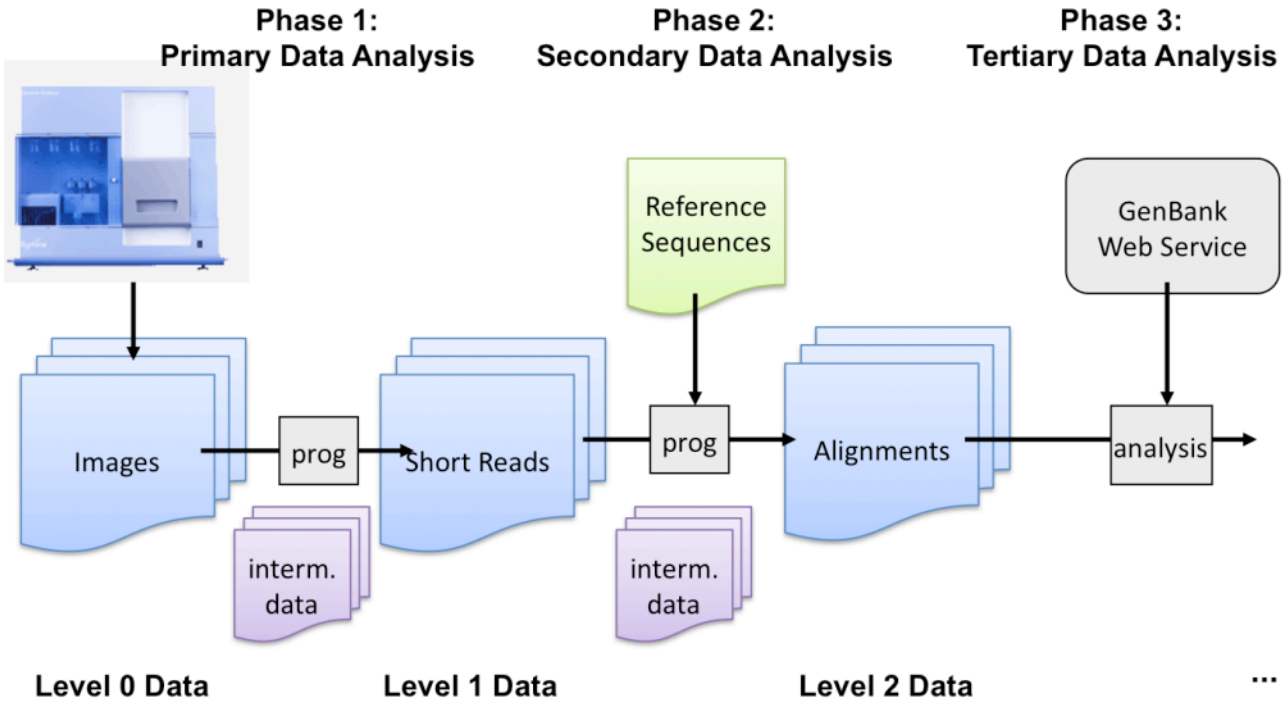

**Phase 0: Sequencer Run**
The content of a flowcell is sequenced in one run of a high-throughput sequencing instrument. E.g., the commonly used Illumina Solexa needs between 2.5 and 5 days for one run [8]. The results are thousands of images as raw level-0 data that show all the base signals in one tile for a certain sequence position.

**Phase 1: Primary Data Analysis**
Next comes an image analysis and base-calling step. The output is level-1 data, a separate file for each lane consisting of millions of short sequence fragments (so-called short-reads) from 35 to 300bp (depending on the technology used). Together with some meta-data about each read (ids, tile, coordinates and quality data) this sums up to several Gigabytes of sequencing data per experiment.

**Phase 2: Secondary Data Analysis**
The next step is to align the small reads to larger sequences. There are two principle approaches: In *de novo* sequencing experiments, the output is unknown and alignment is a computational expensive process of identifying and assembling overlaying fragments to create contigs; in contrast, re-sequencing experiments and digital gene expression studies align the reads against a known set of reference sequences (such as the human genome) to quantify the genes expressed in the samples. The results are ranked lists of (potential) matches as level-2 data.

**Phase 3: Tertiary Data Analysis**
The last step is to qualitatively analyze the alignments, e.g. by comparing the matches of different sequence samples (differential expression analysis); this is based on statistical analysis and also requires access to further information about the matched genes such as their functions or participation in metabolic pathway.

There are several short-read alignment tools available which implement the secondary data analysis phase of re-sequencing / digital gene expression experiments such as MAQ (free/open source), SOAP (free/open source), Eland (part of the Illumina suite), SX Oligo Search (proprietary), and SlimSearch (proprietary) [9]. These tools can be highly parameterized and typically produce several intermediate outputs. For example, MAQ first transforms the output files from a sequencer and the reference sequences into its own internal formats (intermediate binary files); the output of its short-read alignment is another proprietary binary file which then has to be converted into a human readable form before it can be further processed. Note that the final output is a ‘human readable’ text file which actually complicates the further processing in the data analysis step.

### 2.1.1 *Example 1: 1000 Genomes Project*
The 1000 Genomes Project is an international effort to sequence the individual genomes of 1000 humans [1]. One of the participating labs is the Welcome Trust Sanger Institute in Cambridge,UK. The institute has currently 28 Illumina sequencers deployed in a 24x7 production mode. Each sequencer is producing 750 GB of raw level-0 image data per run. After the initial image analysis phase, this results in one short-read file per lane, each of the size between 1 and 2 GB. This totals about 75 TB of raw level-0 data and 0.5 TB of level-1 data per week. This level-1 data is then aligned against the Human Reference genome, resulting in millions of alignments (for quality purposes, individual genomes are sequenced with 40x coverage). The tertiary data analysis phase finally calls the consensus over all alignments, and looks for variations between individual genomes (*single nucleotide polymorphisms* (SNPs)).

### 2.1.2 *Example 2: Digital Gene Expression Studies*
NextGen sequencing is also a promising technique for *digital gene expression* studies [20]. The level of activity of a certain gene is reflected by the amount of messenger RNA (mRNA) found in a cell. State-of-the-art for gene expression studies are Microarrays that only test for a pre-defined set of genes. NextGen sequencing allows performing this analysis *in silico*: mRNA is transformed into DNA fragments, which are then sequenced. The resulting level-1 data, called *tags*, are evidence for gene activities: the more active a gene, the more mRNA was produced, the higher the frequency of the corresponding tag. To identify the genes that are expressed in the cell, the reads are aligned against a reference genome and the alignments are ranked (level-2 data). As tertiary data analysis, one performs a differential expression analysis of different samples, e.g. comparing healthy cells with cancer cells. The data volumes in this second scenario are much smaller than in the previous re-sequencing scenario, because only a faction of the genome is active in a cell and tags are repeating.

**The Role of Databases.** Even though high-throughput sequencing produces massive data volumes and we need to efficiently manage and analyze these data, database systems play only a surprisingly small role. Most data, in fact all level-0 to level-2 data, is stored in a zoo of proprietary file formats in a cluster file system. The only exception is the workflow meta-data (which samples are sampled already to what degree etc) which is stored in a relational DBMS.

## 2.2 Extensible Database Systems
Many DBMSs can be extended with user-defined procedures and functions [12]. In some systems, these programmability facilities are used to support an object-relational data model, but recent developments target the extensibility of the query processing level by providing table-valued functions and even user-defined aggregates. Some DBMSs run their extensions in a hosted runtime such as the JVM or the .Net CLR [2][3].

The normal usage scenario for stored procedures and functions is to implement business transactions that execute with some control flow and logic inside the DBMS engine. Our idea is to leverage such programmability features of modern database engines for implementing advanced gene sequence processing algorithms.

## 2.3 Basic SQL Server 2008 Concepts

An example for an extensible DBMS is Microsoft SQL Server. Since version 2005, it integrates a .NET CLR hosting runtime. The following section briefly describes some basic .NET and CLR hosting concepts that are referenced throughout this paper.

### *2.3.1 .NET Runtime Hosting in SQL Server 2008*

In the .NET framework, many different high-level programming languages can be used to construct programs. The compilation of the program produces a file, called an assembly, containing the compiled code in the Microsoft Intermediate Language (MSIL). The .NET Framework supports a mechanism called custom attributes for annotating classes, properties, functions and methods with additional information or facets the application may want to capture in metadata. SQL Server uses these annotations to complement the specification of extensibility contracts for user-defined functions (UDFs), types (UDTs), and aggregates (UDAs). Managed code is MSIL executed in the CLR rather than directly by the operating system. Managed code applications gain CLR services such as automatic garbage collection, runtime type checking, and security support. These services help provide uniform platform- and language-independent behavior of managed-code applications. At execution time, a just-in-time (JIT) compiler translates the MSIL into native code (e.g., Intel X86 code). During this translation, code must pass a verification process that examines the MSIL and metadata to find out whether the code can be determined to be type safe.

**Figure 2 Integration of the CLR inside MSSQL.**

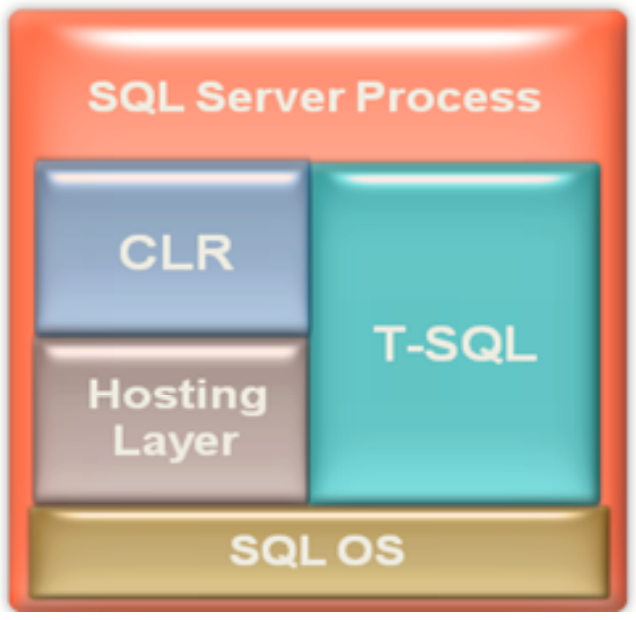

SQL Server and the CLR have different internal models for managing system resources including memory management, synchronization primitives, and thread scheduling. SQL Server acts as the operating system for the CLR when it is hosted inside the SQL Server process (see Figure 2). SQL Server encapsulates all operating system primitives in a component called the SQL OS. The Hosting Layer coordinates assembly loading, security management, application domain management, and escalation policy when critical exceptions occur.

### *2.3.2 User Defined Functions in SQL Server 2008*

CLR UDFs provide a mechanism for extending SQL Server with new scalar and table-valued routines written in .NET languages. Any static method can be registered as a scalar CLR UDF. For example, a new function for a mathematical or statistical operation not natively supported by SQL Server can be written as a static function in C# and registered as a CLR UDF. CLR UDFs also allow SQL Server to leverage the functionality available in the .NET framework such as, e.g, the library for compression (System.IO.Compression).

Table-valued functions (TVFs) provide a mechanism for converting raw data (i.e., in a text/binary file or on the network) into a table. To accomplish this conversion CLR TVFs use a pull model to stream as a collection of rows. SQL Server requests (pulls) and processes one row of data from the CLR TVF at a time as opposed to the TVF materializing the entire relation and then the entire relation being processed. This allows the query engine to begin consuming results immediately after the first row is produced instead of having to wait for the entire return table to be populated. To become a CLR TVF a class implements the IEnumerator interface.

### *2.3.3 User Defined Types in SQL Server 2008*

SQL Server supports UDTs as a mechanism to extend the scalar type system of SQL. SQL Server UDTs, implemented as CLR classes or structs, are designed to model atomic scalar types such as monetary currency, geometry and spatial types, etc. UDTs allow users to define a type using the .NET framework, and to deploy and use these types within the database. Once deployed, UDTs can be used everywhere a native SQL Server scalar type can be used. UDTs can hold state up to 2GB data.

### *2.3.4 User Defined Aggregates in SQL Server 2008*

CLR UDAs give users the ability to write their own aggregates to suite their own specific needs. Some common cases include aggregates for string processing, and statistical or mathematical computations. CLR UDAs are tightly integrated into the system, are used by users just as with native aggregates, and can be parallelized by the system just like built-in aggregates (SUM, COUNT, MIN, MAX). Aggregates can hold state up to 2 GBs and the contract for aggregation accepts multiple columns or expressions as inputs.

### *2.3.5 Page/Row Compression in SQL Server 2008*

SQL Server 2008 also supports row and page compression for tables and indexes. Row compression uses variable-length storage formats for numeric types and fixed-length character strings. Page compression combines row, prefix, and dictionary compression over several rows [11]. The following examples create tables *T1* and *T2* with row and page compression, respectively:

```
CREATE TABLE T1 (c1 int, c2 nvarchar(50))
                WITH (DATA_COMPRESSION = ROW);
CREATE TABLE T2 (c1 int, c2 nvarchar(50))
                WITH (DATA_COMPRESSION = PAGE);
```

### *2.3.6 FileStreams and SQL Server 2008*

The massive data volumes of scientific data management require efficient support of binary large objects (BLOBs), an area for which DBMS are known to not necessarily shine [15]. Microsoft SQL Server 2008 introduced a new functionality of storing BLOB data as files in the file system rather than with the normal relational storage engine. Those *FileStream* BLOBs are under full transactional and managerial control of the database system – all shared access is synchronized via the concurrency control of SQL Server and all file data is managed by the DBMS utilities (e.g. backup/restore, database consistency check). All FileStream BLOBs can be accessed directly using standard Win32 file system APIs from client programs; this access does not go through the database's buffer pool, but rather supports efficient streaming access of the FileStream content to clients [10][14]. As the name suggests, this functionality is mainly targeting efficient access to large BLOBs such as images or videos.

In this paper, we are interested to find out how well such external FileStream BLOBs can be used for storing sequence data.

# 3. DATA MODELING

The common approach to data management for bio-data is to keep data in files of various proprietary formats. Each file format is optimized for a specific task, lacking a common data model. Many formats are optimized for a textual display of the contained data. For example, the common FASTA file format for gene or protein sequences contains line-wrapped sequences to 60 base pairs per line for better readability

**Figure 3 FASTQ File Example**

```
@IL4_855:1:1:954:659
GTTTTTATGGTTTTAGATCTTAAGTCTTTAATCCAA
+
>>>>>>>>>>>>>>>6>>>>>>>;>>>>>>;>>;>;
@IL4_855:1:1:497:759
…
```

Another example is the FASTQ file format (cf. Figure 3) where each entry consists of 4 lines [6]: An entry name, the actual short sequence data, and a comment line. The fourth line is the textual 'printable' representation of so-called *Phred Quality* values per-base. These are the logarithmic-transformed error probabilities from the image analysis phase (value range: 0 to 100) that have been shifted into the visible ASCII character space.

These are good examples of practices that violate the fundamental principle of large-scale data management, data independence.

## 3.1 Conceptual Design

As a first step, we want to develop a concise conceptual data model for gene sequencing experiments that is to be shared throughout all data processing phases. The goal is to identify all data entities and their relationships used in sequencing workflows. Important inputs for this design phase are example files for all workflow phases and a description of the expected analysis results. We used an implementation of the Entity-Relationship data model, the EDM, as a modeling tool [4].

**Figure 4 Conceptual Data Model**

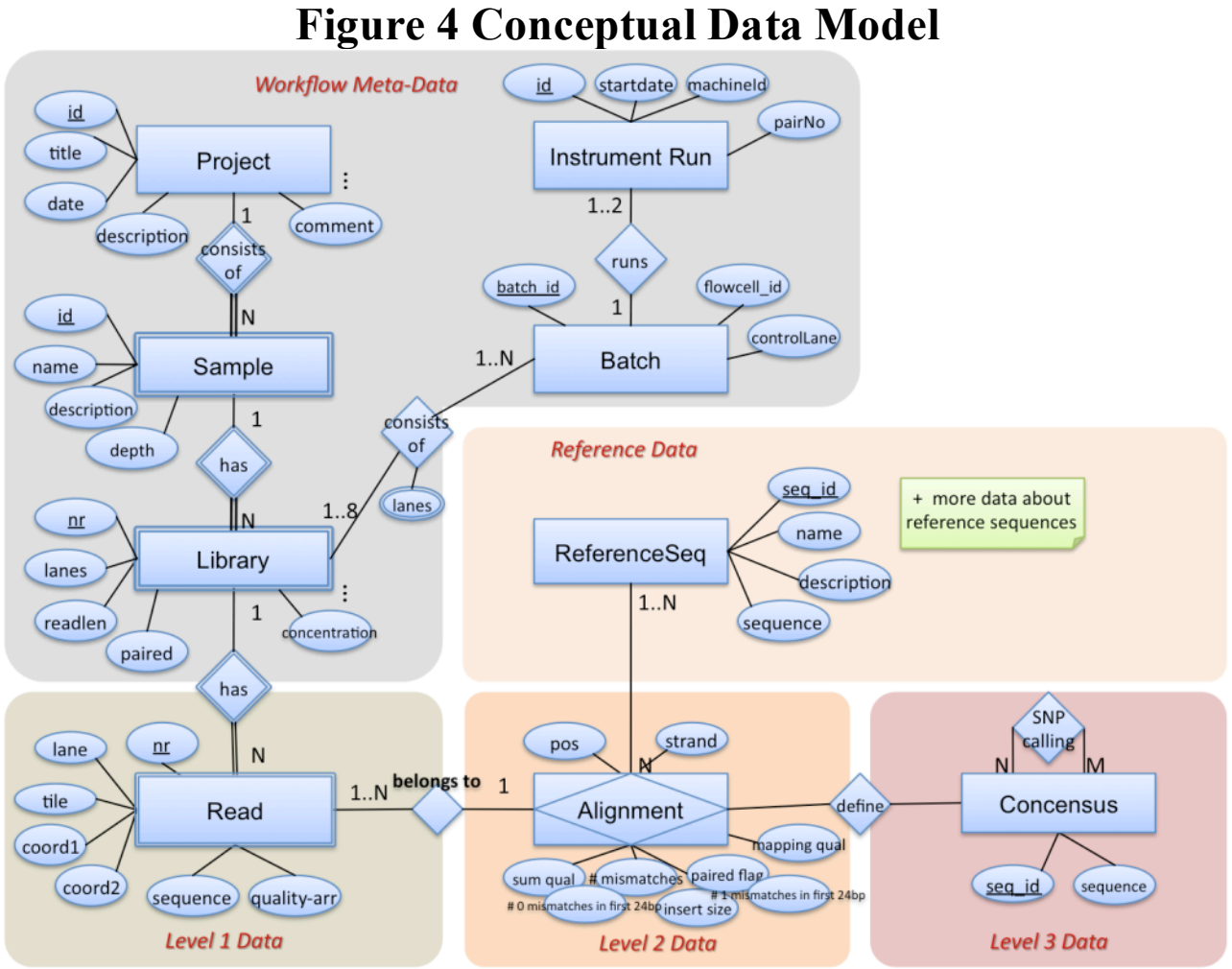

Figure 4 shows the model for high-throughput sequencing starting from level-1 data (due to its size, level-0 data is deleted after the primary data analysis as soon as quality control was performed). There are two challenges for our model: Firstly, the short-read sequences are conceptually probabilistic data that has some error probability values associated with each base (cf. FastQ example above). Secondly, the data model contains both workflow provenance data and the level-1 to level-3 sequence data.

## 3.2 Relational Schema Design

Next, we apply the usual relational mapping rules to our conceptual design to come up with a set of normalized relations for storing the sequence data. Although a natural step in relational database designs, the outcome is a drastic departure from the state-of-the-art in file-centric bio-data management – both workflow management data and sequencing data are now integrated in one schema. This allows scientists to explore the context of their experimental results with a simple navigational query. Furthermore, a normalized relational schema strives to minimize redundancies. Instead of repeating the short-read sequences in each relation, they are linked back to the base relation with the level-1 data by foreign-key relationships.

**Example.** In the 1000 Genomes Project, the result of the secondary data analysis is a set of alignments of the short-read sequences from one experiment against the Human reference genome. With a file-centric data management, the alignments file actually lists each aligned short-read with the id and position of the reference sequence it has been aligned to. In our normalized schema, this is represented by the `Alignment` table that links with a foreign key back to the actual short-read in the `Read` table.

## 3.3 Physical Database Design

We finally have several possibilities on how to organize an efficient physical database design for our sequencing data:

A full relational approach imports all data into the relations and makes the relational database engine solely responsible for all data management. Such a **database-centric approach** has all the advantages of central data management with a database system such as transactions, declarative query processing with automatic query optimization by the DBMS, controlled shared access to the database, and backup/recovery facilities. But at the same time, the DBMS becomes responsible to process the sequence data such as sequence alignments and base calling. The alternative of running those data analysis steps outside the DBMS would otherwise mean to export the relational data into the proprietary file formats expected by the analysis tools. This would render the intended central data management with a database system ad absurdum.

In contrast, a **hybrid approach** keeps (some) base data in files and accesses them through a wrapping mechanism in a relational view in the DBMS. The benefit is that existing bioinformatics tools can be used almost unchanged: they are allowed to process their files as expected and even with the same file-scan and seek API calls as before. Only their file opening code would have to be adapted as access to the files is now managed by a DBMS. In addition, we can query the base data in the database, combine (join) it with other relations and also make use of the database's access control and backup facilities.

**Example.** We demonstrate the hybrid approach using the FileStream feature of SQL Server 2008: We want to store short-read sequences of a typical sequencing workflow in their original FASTQ format, while at the same time being able to access them via SQL queries. This requires importing the data into the relational schema as a special BLOB that is stored as FileStream and then accessing the sequence data through a relational wrapper table-valued function (TVF). The following Transact-SQL (T-SQL) statements create a corresponding `ShortReadFiles` table, import the FASTQ file (in this brief example, we just bulk-import the data; note that external tools can also write the file content using the normal WriteFile() API calls when needed) and list the content using a `ListShortReads()` TVF function:

```
CREATE TABLE ShortReadFiles (
   guid   uniqueidentifier ROWGUIDCOL PRIMARY KEY,
   sample  INT,
   lane      INT,
   reads     VARBINARY(MAX) FILESTREAM
) FILESTREAM_ON FILSTREAMGROUP;

/* Bulk-Import new FileStream row */
INSERT INTO ShortReadFiles (guid, sample, lane, reads)
 SELECT NEWID(), 855, 1, *
 FROM OPENROWSET(BULK'D:\855_s_1.fastq',SINGLE_BLOB);

/* check meta-data of the filestream table content */
SELECT guid, ample,lane,reads.PathName(),DATALENGTH(reads)
  FROM ShortReadFiles;

/* check content of one FileStream column using a TVF */
SELECT *  FROM ListShortReads( 855, 1, 'FastQ');
```

# 4. DATA PROCESSING

There are many specialized bioinformatics algorithms for processing and analyzing sequencing data. If databases shall have a more prominent role for bio-data management, they have to support such functionality too – either as embedded functions or by integrating with the existing file-based programs or Perl scripts such as MAQ or NCBI Blast. In the following, we study on how we can achieve efficient data access to FileStream data inside the database, and how this can be leveraged to implement some analysis steps directly with SQL.

## 4.1 File Wrapper Functions

The main challenge with the hybrid approach is to be able to efficiently access the file content from the relational query engine. To be able to do so, we need to provide efficient set-valued file wrapper functions. Given the large data sizes of today's genomics (file sizes of 500 MB and more are not unusual) an efficient, streaming-like access is important.

Most database engines allow extending the core database functionality with user-defined functions and stored procedures. In SQL Server, one can write user-defined functions in either T-SQL, or in any .NET language. User-defined functions can be either scalar or table-valued. The API for providing TVFs follows the standard iterator interface of a relational query engine [17]. This means that programming a table-valued function is much closer to implementing a user-defined relational operator than writing a schema-level UDT method.

Figure 5 above shows a sequence diagram of the interaction between the query processor (QP) and a table-valued file wrapper function (TVF) that parses data stored in a FileStream BLOB. The actual TVF does not do much but rather all data access (and file parsing) logic concentrates in the TVF's result iterator. The query processor uses this iterator to scan through the result set and calls for each row an explicit data conversion function (FillRow) that converts the internal CLR data into the equivalent SQL data types.

Given this iterator-oriented contract between the query engine and the file wrapper function, how can we provide fast streaming access to data stored in a FileStream?

**Figure 5 Sequencing Diagram of File Wrapper TVF**

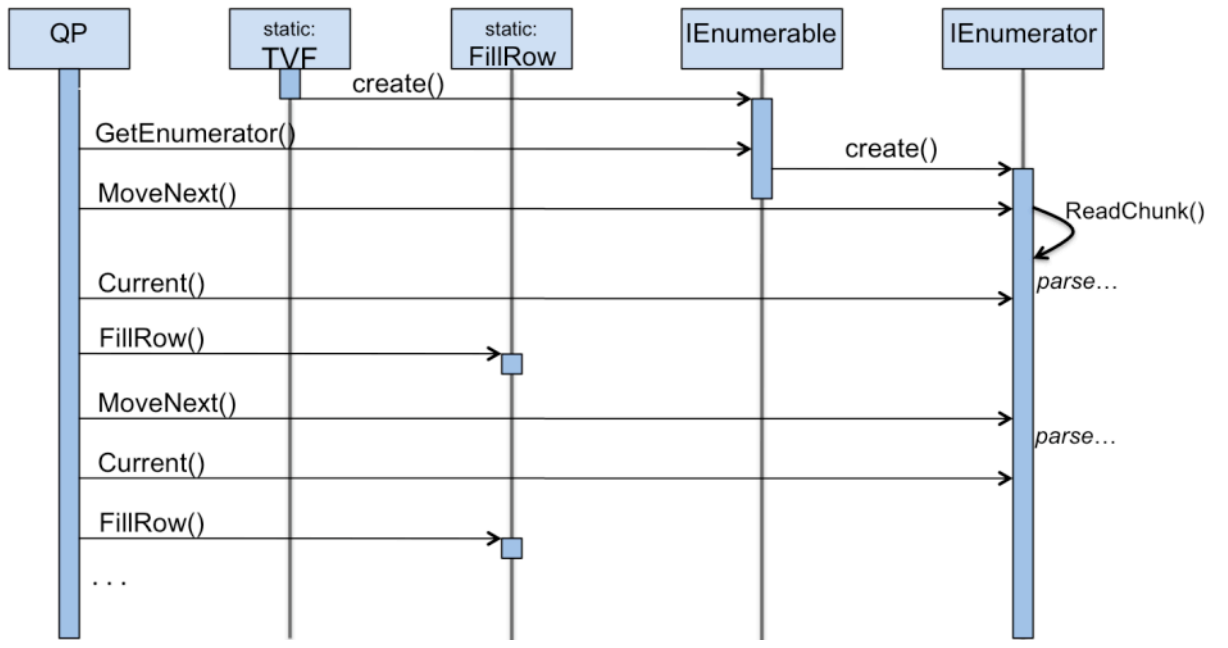

The solution is to separate the file parsing logic from the file-streaming. Instead of reading individual lines form a text file with each `MoveNext()` call, we wrote a paging function that scans through the file in larger chunks of data (cf. the `ReadChunk`() call in Figure 5). The parsing code works on an internal byte buffer of the iterator that has been filled by the previous `ReadChunk`() call. An additional buffer-wrap check is needed in the parsing code, because the chunks are not guaranteed to be multiple of file entries; if the end of the chunk is reached before the parsing of the current entry is finished, a paging algorithm moves the current incomplete entry to the start of the buffer and fills the remaining buffer with the next chunk. Finally, we can stream fast through a FileStream BLOB by using the `GetBytes` function together with the `SequentialAccess` flag that implements pre-fetching on FileStream data. The following code gives an overview of our table-valued FileStream wrapper:

```
Iterator::Create():
  filePos = 0; bufferPos = 0;  bufferOffset = 0;
  bufferSize = buffer.length();   bytesRead = ReadChunk();

Iterator::MoveNext():
  while ( bytesRead > 0 ) do
      if  (bufferPos >= bytesRead) then
          bytesRead = ReadChunk();
      fi
      pos = ParseShortReadEntry(buffer, bytesRead, bufferPos);
      if ( pos > 0 ) then
          bufferPos += pos;
          return true;
      else // paging algorithm
          MemCopy(buffer, bufferPos, 0, bytesRead-bufferPos);
          bufferOffset = bytesRead-bufferPos;
          bufferPos = bytesRead; // will trigger next ReadChunk()
      fi
  done
  return false; // end-of-file reached

Iterator::ReadChunk():
  len   = bufferSize – bufferOffset;
  read = sqlinput.GetBytes(0, filePos, buffer, bufferOffset, len);
  filePos += read;  bufferPos = 0;
   if ( read > 0) and (bufferOffset > 0) then
      read += bufferOffset;
      bufferOffset = 0;
  fi
  return read;
Iterator::ParseShortReadEntry(…):
 // encapsulates file parsing logic – returns 0 if buffer end reached
```

## 4.2 Typical Genomic Queries

One of the main benefits of a database-centric (and also the hybrid) approach is that we can use declarative SQL queries to analyze the bio-data. The query processor then can provide us with automatic parallelization of the data access.

In the following, we just give a few examples for how data access to relationally stored bio-data can be done using SQL queries.

### 4.2.1 *Example 1: Unique Short-Reads*

One of the first steps of a digital gene expression study [20] would be to rank the level-1 sequence results by their frequency: How often were certain sequence tags found in the sample (ignoring all uncertain sequences, i.e. such which contain at least one 'N' symbol)? The corresponding short-read analysis (sometimes referred to as *binning*) is a typical scripting problem where in one example a 26-line Perl script was found to perform this function. The same functionality can be achieved with an aggregation-group-by query as follows:

**Query 1 - Binning Unique Short Reads**

```
SELECT ROW_NUMBER()
       OVER (ORDER BY COUNT(*)DESC),
       COUNT(*),
       short_read_seq
 FROM [Read]
WHERE r_e_id=1 AND r_sg_id=2 AND r_s_id=1
      AND CHARINDEX('N',short_read_seq)=0
GROUP BY short_read_seq
```

Note how the filtering for short-reads without 'N' symbol becomes a straight forward WHERE predicate. The main challenge is to determine a rank number for each tag. In our query, we use a proprietary extension of T-SQL, the ROW_NUMBER() function, to determine the result row number.

### 4.2.2 *Example 2: Digital Gene Expression Analysis*

The following is an example of a tertiary data analysis: We would like to determine which genes are expressed in the short-reads found in a certain sample of a digital gene expression experiment. We hence have to analyse against which reference sequences the short reads have been aligned to and how often. Again, this is a classical example of a scripting problem, which given our relational approach can be expressed by a simple grouping query:

**Query 2 - Gene Expression Analysis**

```
INSERT INTO GeneExpression
 SELECT  a_g_id, a_e_id, a_sg_id, a_s_id,
         SUM(t_frequency), COUNT(a_t_id)
   FROM Alignment JOIN Tag ON (condition)
  WHERE a_e_id=1 AND a_sg_id=1 AND a_s_id=1
  GROUP BY a_g_id, a_e_id, a_sg_id, a_s_id
```

This query groups all aligned tags by the gene id (a_g_id) and determines the total frequency and count of the tags aligned against the same gene. Most of the query's complexity is due to the use of composite primary keys. For example, each tag is associated to a sample of a specific experiment (a_e_id, a_sg_id, a_s_id). This is a result of the normalization of the schema. It however also allows to easily select alignments from a specific experiment or sample.

### 4.2.3 *Example 3: Consensus Calling*

Next, we consider an example for a tertiary data analysis that allows us to make good use of the extensibility features of modern DBMS. The *consensus calling* step of a re-sequencing project, such as the 1000 genomes project, aggregates all alignments for the same sample to one result sequence (e.g. the one individual genome of a person that has been sequenced multiple times). The alignments overlap and the algorithm has to decide on the resulting base symbol for each position of the result sequence. Figure 6 illustrates this process.

**Figure 6 - Consensus Calling Overview**

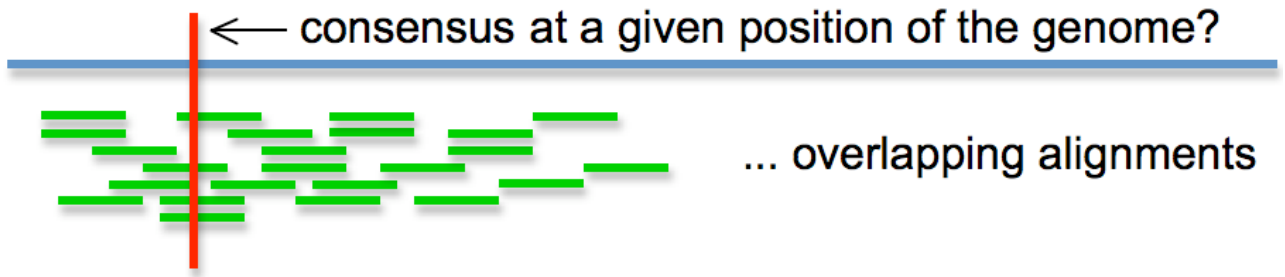

In an extensible database system, we can solve this task in a declarative way with one SQL query and a combination of several TVFs and UDAs. A conceptually clean approach is as follows:

**Query 3 - Consensus Calling in SQL**

```
SELECT chromosome, AssembleSequence(pos,b)
  FROM(SELECT chromosome,pos,CallBase(base,qual) b
         FROM Alignments JOIN [Read] ON (…)
         CROSS APPLY PivotAlignment(pos,seq,quals)
        WHERE a_e_id=…  // join condition
        GROUP BY chromosome, pos )
 GROUP BY chromosome
```

This is a query where the CLR integration of a database engine can shine. The core idea behind Query 3 is to decompose the consensus calling algorithm into small building blocks, which are then naturally combined in a declarative query:

`PivotAlignment(pos, seq, quals)` is a table-valued function that transforms a given DNA sequence with quality values (seq, quals), that is aligned against a reference sequence at position (pos), into a list of (position, base, qual) tuples.

`CallBase(base,qual)` is a user-defined aggregate function that aggregates all bases that have been aligned against the same position to the consensus base. This base calling can take the quality values of each aligned short read into account too.

`AssembleSequence()` is another user-defined aggregate that does the reverse of PivotAlignment(), namely concatenating all called bases to the resulting consensus sequence.

Query 3 first pivots all aligned short reads into a long list of *(position, base)* pairs that represent the individual bases aligned against the reference sequence. Next, it groups those individual bases per position and chromosome. The actual base calling is then done as an aggregation step on these groups. Finally, it assembles the resulting consensus sequence – again as an aggregation over all groups of the same chromosome.

However, this approach will generate a large intermediate result when pivoting each alignment into its individual bases and then grouping those by position and chromosome. An optimisation is hence to combine base-calling and result sequence assembly in one `AssembleConsensus()` UDA that combines all overlapping alignments for a given chromosome at once without pivoting. To avoid constructing the large intermediate result as internal state of that function, we propose a sliding window processing technique and to scan over the alignments in order of their starting position.

## 5. EVALUATION

We prototyped two sequencing scenarios: (a) digital gene expression studies, and (b) the 1000 Genomes project using the proposed hybrid data management approach on SQL Server 2008. In the following, we present some initial results on the quality of our approach in terms of processing and storage overheads. We concentrate on the most significant results. The test environment is a Dell OptiPlex 755 (2 Intel Q6600 dual core CPUs at 2.4 GHz, 8 GB DDR2 RAM, two 250GB SATA disks) under Windows Server 2008 Enterprise.

### 5.1 Data Size Comparison

First, we are interested in storage efficiency of the different data management approaches. We compare the storage requirements of the original separate input files (Files), with the hybrid FileStream approach, and several full-relational approaches. The example data sets are two complete sets of level-1, level-2 and level-3 data from a single flowcell lane of two different high-throughput sequencing experiments.

#### *5.1.1 Digital Gene Expression Data*

In Table 1, we compare the amounts of data produced by a digital gene expression study: a 490 MB short-reads file for one lane, a 16 MB output file from a unique tag analysis, 45.8 MB alignments data (output of a run of MAQ against the Human reference genome) and a 1.8 MB result file of a gene expression analysis of the alignments.

**Table 1 Comparison of Storage Efficiency (Digital Gene Expr)**

| | | | | Normalized Schema | | |
|---|---|---|---|---|---|---|
| | Files | FileStream | 1:1 Import | plain | row comp. | page comp. |
| Short Reads | 490.0 MB 5,028,052 rows | 490.0 MB | 965.3 MB | 493.7 MB | 467.0 MB | 235.8 MB |
| Unique Tags | 16.1 MB 565,526 rows | n/a | 23.4 MB | 24.2 MB | 19.8 MB | 17.4 MB |
| Alignments | 45.6 MB 563,985 rows | n/a | 58.3 MB | 37.9 MB | 26.6 MB | 12.5 MB |
| Gene Expressions | 1.8 MB 24,951 rows | n/a | 2.3 MB | 712 kB | 392 kB | 328 kB |
| Total | 553.5 MB | | 1049.3MB | 556.51MB | 513.8 MB | 266 MB |

As we can see, FileStreams have no storage overhead as compared to the original files. The data is just stored in its original format and size on a FileStream as on a file on disk. For the purely relational representations, the results are more diverse.

If we import all sequence data files unchanged (1:1) from the source files, the storage size almost doubles. Note that this is not the proposed normalized schema, but a simulation of a user trying to use a relational database in a 'straightforward' manner just based on the input file formats. The reason is twofold. First, there is a slight overhead with the internal row structure in the storage engine. Second, the main reason for the overhead is redundancies with id attributes. Since there is no central ID management in a file-centric storage or schema information to declare key attributes, the common approach is to construct for each record a unique name by combining several attribute values. In other words, the approach is to materialize a composite primary key.

For example, the name of an individual short read entry in a FASTQ file is a string that combines the name of the sequencer machine with the flowcell id, the lane and tile numbers on the flowcell, and the x and y coordinates on the tile where that sequence has been located. If one imports the data files with those textual IDs into a relational schema that mimics the file structures, thereby repeating those textual identifiers in several tables, then the storage space increases massively.

The solution is to normalize the schema and to introduce synthetic unique ID values for each database entry. In a plain normalized relational schema we achieve the same storage efficiency as with the original files. The more compact IDs and the minimized redundancy compensate for any overhead of the relational tuple storage. We can further improve this by using transparent row- or page-level compression in the tables [11]. This experiment illustrates that the storage representation in a database approach compares well with a file-centric approach.

#### *5.1.2 Data from the 1000 Genomes Project*

We also studied the storage efficiency of the different approaches with data from a re-sequencing experiment. Table 2 compares the storage requirements of the different processing stages of one lane short-reads data from the 1000 Genomes project: ca. 6.2 million short reads, which are aligned against the 25 chromosomes of the Human reference genome.

**Table 2 Comparison of Storage Efficiency (1000 Genomes)**

| | | | | Normalized Schema | | |
|---|---|---|---|---|---|---|
| | Files | FileStream | 1:1 Import | plain | row comp. | page comp. |
| Short Reads | 607 MB 6,271,727 rows | 607 MB | 1072 MB | 656 MB | 627 MB | 574 MB |
| Reference Seq | 2910 MB 25 rows | 2910 MB | 2999 MB | 2999 MB | 2999 MB | 2999 MB |
| Alignments | 1516 MB 11,418,757 rows | 1516 MB | 1528 MB | 904 MB | 787 MB | 745 MB |
| Total | 5033 MB | 5033 MB | 5599 MB | 4559 MB | 4413 MB | 4318 MB |

The characteristics of this data set are different than for digital gene expression studies: Almost all short reads are unique which leads to an order of magnitude larger number of alignments. Still, the results in principle support our previous findings. FileStreams have no storage overhead as compared to the original data files. A simple 1:1 import of the data that keeps its textual identifiers is actually larger than the original data sizes, while a normalized schema using numerical identifiers and foreign keys is much smaller (e.g. for the alignments, we can save 40% space this way).

However, the database internal compression on page and row level is much less effective on this data set than in the previous scenario. The reason is that the short-reads are much less uniform and hence the common-prefix- and dictionary-based compression algorithms over only a small subset of the data fitting on one disk page do not perform that well. In this case, larger chunking or a different compression algorithm would be beneficial (e.g. a bit-encoding of the sequences could reduce the size to just about a quarter). This could be achieved by introducing a corresponding domain-specific short-read data type.

### 5.2 File Wrapping Performance

Next, we are interested in the performance of different approaches to access short-read data depending on the physical design. In particular, we want to compare the plain file access with a purely relational access and different approaches to scan FileStream data.

In the following scenario, we execute a `SELECT COUNT(*) FROM File` query with those varying functions. The example data set is a short-reads file in FASTA format with 5,028,052 lines of short read data from one lane of a sequencing experiment. Note that in this experiment, the function did *not* perform any record conversions.

| | |
|---|---|
| Command line program (C#) | ~ 5 secs |
| T-SQL Stored Procedure | several minutes |
| CLR-based Stored Procedure with StreamReader | 21 secs |
| CLR-based Stored Procedure with Chunking | 7 secs |
| CLR-based TVF with Chunking | 14 secs |

These results demonstrate that CLR-based stored procedures and user defined functions can provide performance comparable to file-based tools. TVFs are slower than stored procedures due to the overhead of the iterator-interface logic – but they provide a streaming access to the data, while stored procedures materialize their set-valued result [17].

The biggest performance bottleneck is the actual data conversion of FileStream records into SQL tuples, which is done by the FillRow() function in Figure 5. This requires not only parsing of the original data, but also copying each attribute value from the FileStream buffer into a corresponding result tuple. As the FileStream buffer resides inside the CLR, the data has also to be copied and converted from the CLR sandbox into the SQL query engine. This adds a large overhead on the pure streaming performance as shown above.

## 5.3 Data Analysis Capabilities

Let us finally consider how well database systems support the different data analysis phases of a typical genomics workflow:

### *5.3.1 Primary Data Analysis*

Our approach starts from level-1 data – the short read sequences. Hence the image analysis of the raw image data still has to be file-centric. The only support needed is that the tools write the resulting level-1 data into FileStreams managed by the DBMS. This approach is consistent with the current practice of sequencing labs that keep level-0 data only for a few weeks for quality control purposes, before deleting it.

Recently, the *Sequence Read Format* (SRF) was proposed as a common container format for level-1 short-read data [16]. SRF files include not only the actual short reads and quality values, but also some core information from the image analysis steps such as intensity and signal-to-noise ratio values. In our experiments, we did concentrate on the currently still more common FastQ format. Our hybrid approach would however naturally extend to encapsulate SRF files as FileStreams too.

### *5.3.2 Secondary Data Analysis*

The core idea of the hybrid approach is to leverage on the existing tools and algorithms for secondary data analysis such as MAQ by allowing them to still work on the same input files. This requires opening files through the new FileStream-aware file system APIs. The remaining access and access speed is unchanged. Alternatively, we can implement the alignment algorithms directly in the DBMS as stored procedures. Previous work showed that this is possible, although with limited scalability [13].

Some secondary data analysis can be implemented solely in SQL, such as some preparation steps for digital gene expression studies. In Section 4.2.1, we were discussing a declarative way for binning unique short reads as preparation step for the actual sequence alignment. We compared the runtime performance of Query 1 versus an equivalent Perl script, which we found used by some bioinformatics colleagues, over a 500 MB short-read data set. Both approaches where producing the same 565,526 unique reads. The Perl script ran for 10 mins, while SQL Query 1 on SQL Server 2008 finished in just 44 sec.

**Figure 7 - Resource Consumption of Perl Script**

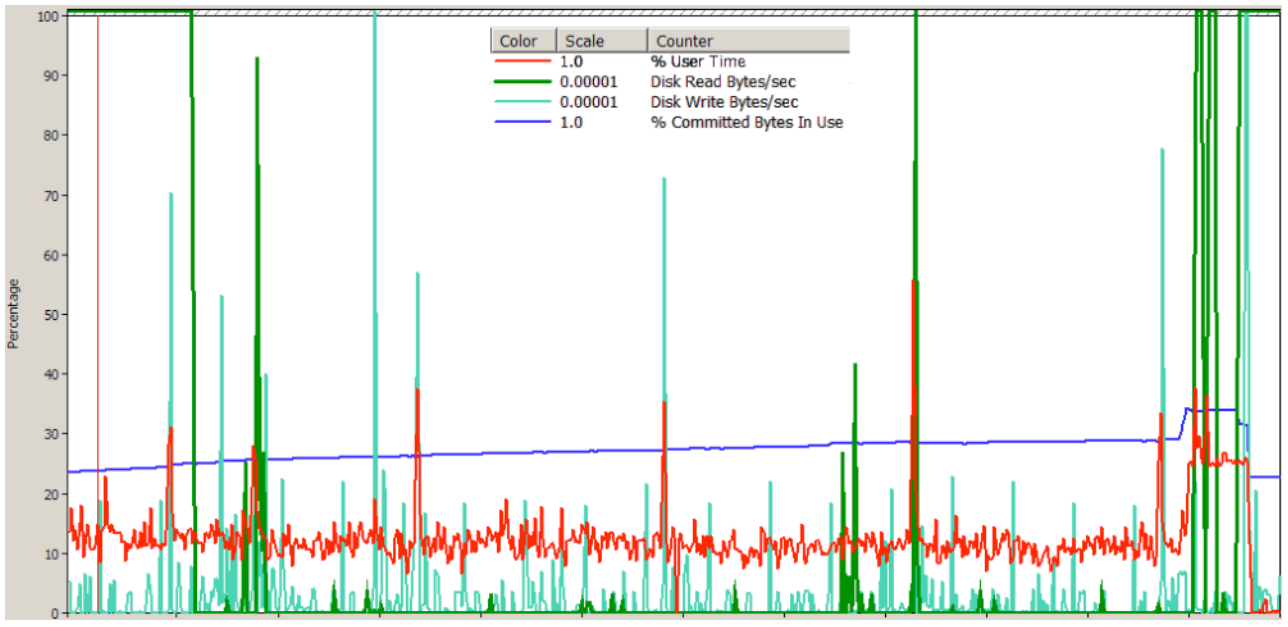

The main reason is that the Perl script follows a typical sequential scripting approach to bio-data processing. As we see in Figure 8, it first reads all data into main memory (dark green line – CPU read activity), before it then processes the data until finally writing the result back to disk. All throughout this processing, only about a quarter of the available CPU power is used (dark red line) because it runs sequentially on just on of the four available cores. In contrast, SQL Server automatically parallelized the query and made use of all four cores, resulting in a much faster runtime (as shown in Figure 8 and Figure 9 below).

**Figure 8 - Multi-Core-CPU Usage of SQL Query 1**

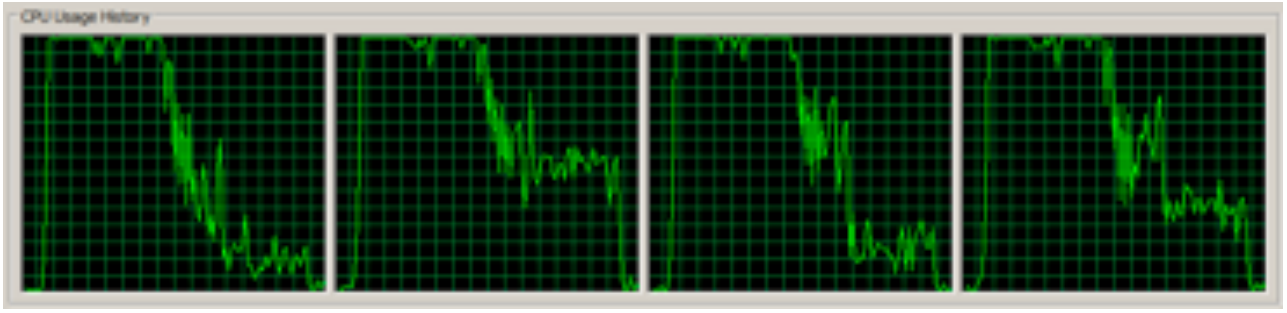

### *5.3.3 Tertiary Data Analysis*

The most interesting case for using a database engine for genomic data management is the tertiary data analysis phase.

In Section 4.2.3, we discussed a way for implementing the consensus calling step of a sequencing project in one SQL query using some user-defined aggregate functions. We conducted some initial experiments with different versions of Query 3 on a data set of 2 GB consensus data from just a single lane of human genome sequence data. The results again show how the query optimizer gives us parallelism for free, but they also highlight some current limitations for sequence processing in databases.

**Figure 9 - Parallel Query Plan for Unique-Read Binning in SQL (Query 1)**

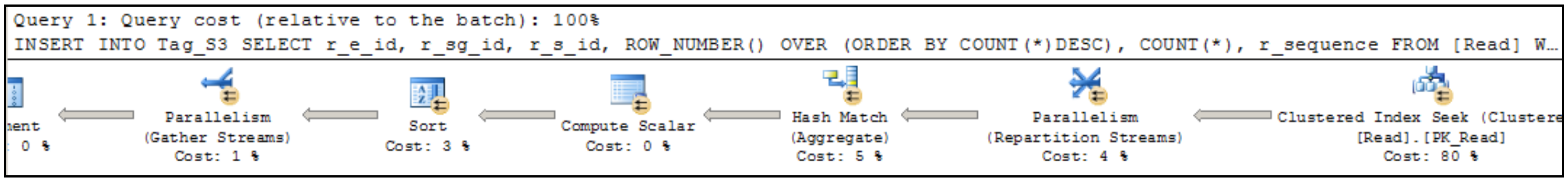

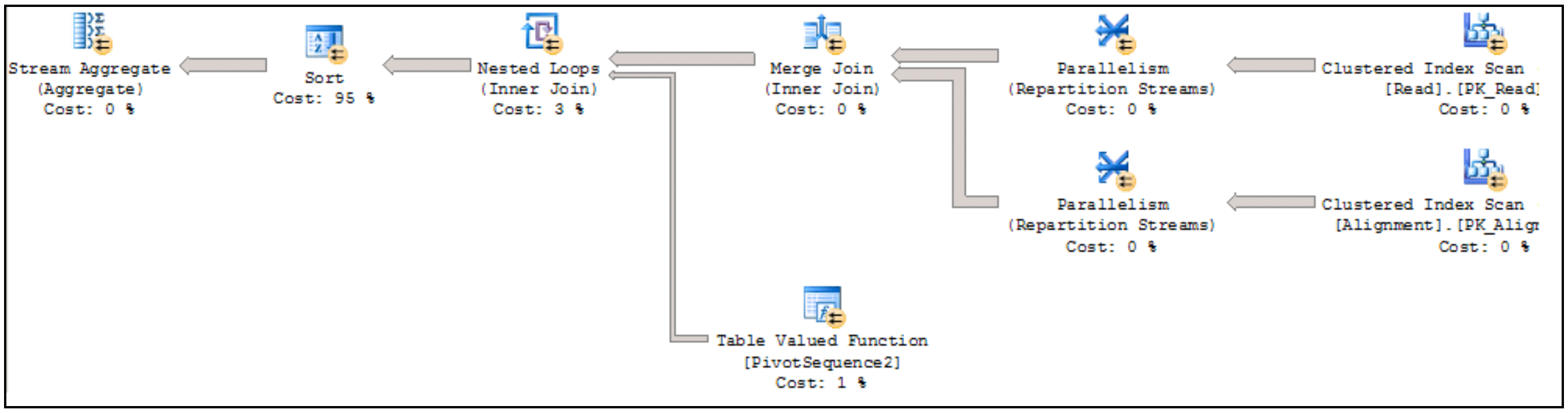

**Figure 10 - Parallel Query Plan for Consensus Building in SQL (Query 3; excerpt)**

Figure 10 above shows the query execution plan for Query 3 (note: due to lack of space, only an excerpt with the core parallel join is shown – the remaining upper execution plan, with the second parallelized aggregate and the projection, is truncated on the left). This query runs on the normalized database schema which stores short reads centrally in a Read table and links to the remaining information via foreign keys. The query hence first has to join the alignments with the short read tables to retrieve the sequences for each alignment. Due to physical data independence, we can choose appropriate clustered indexes on those tables so that the query processor can do this join in about 7 seconds (with a warm buffer pool) by using a parallel merge join. This corresponds to about 1.6 million alignments per second.

The overall performance is dominated by the next processing step. If we pivot each aligned sequence using the *PivotSequence* user-defined TVF as discussed in Section 4.2.3, we produce a huge intermediate result on the temporary tablespace, which we then have to group for the base call and sequence assembly steps. Although conceptually clean, this approach means blocking operations in the execution path and large amounts of disk writes for the intermediate results. Hence it is not practical.

For fastest performance, the database needs to use a non-blocking, parallelized query plan and to processes the alignments in order. When we scan once over all alignments in ascending order of alignment positions, we can develop a UDA that builds the consensus sequence in a sliding-window fashion. Two issues remain: The query optimizer has to find a partitioning of the input data that allows for parallel processing of separate alignment streams even though alignments on the partition borders overlap the next partition – and hence need to be considered in both. Secondly, the final result is one consensus sequence per chromosome – which in the case of the Human genome means a resulting sequence string of more than 100 MBytes.

Consensus calling hence needs an approach to efficiently handle a large internal BLOB result, such as a specific sequence data type that supports streaming inside a database engine. A more thorough evaluation of this tertiary data analysis step is planned, but we are currently lacking the required full set of alignment data from several DNA samples of the same individual for this experiment.

# 6. SUMMARY

In this paper, we outlined an approach to data management for high-throughput sequencing workflows by leveraging the data modeling, storage, query, and extensibility tools of relational DBMSs. We developed an E-R model for genomic data and explored different physical data representations for it, including a novel hybrid design that wraps level-1 sequence data as FileStreams so that existing tools can still be used. Based on this, we demonstrated how the integration of modern programmability features with declarative SQL queries, such as CLR-based user-defined functions and aggregates, can be used for secondary and tertiary data analysis.

Our experimental results show that a normalized database in combination with the FileStream BLOB storage can reduce the storage requirements as compared to purely file-based approaches. We also demonstrated how declarative querying of biological data can be concise, flexible, and how a database query processor can provide transparent parallel processing – given the right physical design choices. We also identified some performance bottlenecks due to missing native data types and the difficulties for the query optimizer to determine the selectivity and the optimal parallelization steps for UDTs and UDAs.

Overall, we believe a database-centric approach can bring many data modeling and performance advantages to the management of high-throughput genomic data.

## 6.1 Future Work

There are several interesting directions for future work: We have seen that the bioinformatics area can benefit a lot from conceptual modeling. But more work needs to be done to successfully propagate and apply the principles of good data design and physical data independence needs to the scientific community.

Our case study so far did just scratch the surface of the actual secondary data analysis stage that requires efficient alignment algorithms. It would be interesting to investigate how a set of core functions and alignment algorithms such as MAQ could be integrated into database systems. There were already some approaches to do so for the popular BLAST alignment algorithm [18]. But none has looked into multi-sequence alignments for high-throughput sequencing yet. We also did not study appropriate indexing. Previous studies, e.g. on integrating BLAST into database systems, showed though that efficient indexing plays a key role in performance optimizations for sequence alignment in databases [7][13].

Our investigation of the storage requirements demonstrated that it would be beneficial to introduce specific genomic sequence data types that support internal compression to reduce the storage requirements for large-scale genomic experiments. Furthermore, one of the benefits of declarative querying over large amounts of data is the possibility to introduce automatic parallelisation.

However, our performance results in Section 5 showed that there is room for improvement on how to efficiently parallelise user-defined functions and aggregates. We are currently looking into these technical aspects and further performance improvements.

Besides those system engineering aspects, there are also many interesting problems on the data model side. Short-read sequence data is probabilistic data as represented by the quality values associated with each read. However, so far many algorithms simply ignore those quality values, and also our own conceptual model does not integrate these aspects well, resulting in separate attributes for sequence data and quality values. An approach with probabilistic databases hence seems natural.

A final important, also quality-related aspect is data provenance management: When and how were short-reads sequenced, which alignment algorithm with certain parameters was used to align them against (a specific version of) the Human reference genome? These are central questions to control the quality of sequencing results and to identify potential problems or missing data. Our current approach does not cover these data provenance aspects. Recently, there is a lot of interest in the database community on data provenance management and it will be interesting to see how those approaches can be adapted for high-throughput genomics.

## Acknowledgements

We are grateful to Matt Wood and Roger Pettett from the Sanger Institute, UK, for their support, as well as to Todd Smith and Eric Olson from Geospiza, Inc., Seattle, for many fruitful discussions.